\documentclass[11pt,a4paper]{article}
\usepackage{jcappub}

\newcommand{\pd}{\partial}

\newcommand{\diag}{\mathop{\mathrm{diag}}\nolimits}

\newcommand{\Fc}{\mathcal{F}}
\newcommand{\Gc}{\mathcal{G}}
\newcommand{\Mc}{\mathcal{M}}

\newcommand{\Lc}{\mathcal{L}}

\title{Perturbative stability of SFT-based cosmological
models}
\author[1]{Federico~Galli%
\note{Aspirant FWO-Vlaanderen.}}
\author[2]{and Alexey~S.~Koshelev%
\note{Postdoctoral researcher of FWO-Vlaanderen.}}
\affiliation{Theoretische Natuurkunde, Vrije Universiteit Brussel and\\
The International Solvay Institutes,\\
Pleinlaan 2, B-1050 Brussels, Belgium}

\emailAdd{fgalli@tena4.vub.ac.be}
 \emailAdd{alexey.koshelev@vub.ac.be}

\abstract{
We review the appearance of multiple scalar fields in linearized SFT based
cosmological models with a single non-local scalar field. Some
of these local fields are canonical real scalar fields and some are
complex
fields with unusual coupling. These systems only admit numerical or
approximate 
analysis. We introduce a modified potential for multiple scalar fields 
that makes the system exactly solvable in the
cosmological context of Friedmann equations and at the same time preserves the
asymptotic behavior expected from SFT. The main part of the paper consists of the
analysis of inhomogeneous cosmological
perturbations in this system. We show numerically that perturbations
corresponding to
the new type of complex fields always vanish. As an example of application of
this model we consider an explicit construction of the phantom divide
crossing and prove the perturbative stability of this process at the linear order.  The issue of ghosts and ways to resolve
it are briefly discussed.}

\begin{document}
\maketitle
 
 \flushbottom


\section{Introduction}\label{intro}

A new promising class of cosmological models originated from  string field
theory (SFT) \cite{review-sft} and 
$p$-adic string theory \cite{padic} has emerged recently and attracts nowadays
a lot of attention \cite{IA1,nongauss,Calcagni,Cline1,Biswas_et_all,KVlast,Koshelev:2007fi}. The main distinguishing feature of
such models is the presence of specific  non-local operators in the action.
There is a strong theoretical reason to consider such non-local
interactions. 
In order to have a good effective model the UV-completeness of the underlying theory is essential  
and these non-local operators are crucial for  making  SFT and $p$-adic string theory UV-complete \cite{review-sft,padic}.
Furthermore, such non-local operators have proven to give rise to models with new appealing properties 
in the context of cosmology,\footnote{Other non-local models were considered as well in
the literature \cite{non-local-set} in the cosmological context.} and
various interesting results have been obtained recently.

In application to
the Dark Energy (DE) problem,
one can easily construct models obeying the phantom divide crossing.
This is of  interest for the present cosmology. Indeed  recent results of WMAP \cite{Komatsu:2008hk}
together with the data on Ia supernovae and  galaxy clusters
measurements, give the following bounds for the DE state parameter $
w_{\mathrm{DE}}=-1.02^{+0.14}_{-0.16} $.  The  crossing of
the phantom divide as well as the phase with $w<-1$ are not excluded
experimentally. Moreover, the experimental data for cosmological epoch corresponding to redshift $z>1$ leave even more freedom for
possible values of $w$ during that phase.

The phantom divide line $w=-1$, corresponding to pure vacuum energy,  gives the lower bound
of the Null Energy Condition (NEC), whose physical motivation is to prevent instability of the vacuum.  Therefore
in order to obtain a stable model with
$w<-1$ one should construct an effective theory, with the NEC  violation,
from a fundamental theory which is stable and admits quantization.
This is a hint towards  SFT inspired cosmological models.  It is therefore important to  find models where the DE
crosses  the phantom divide.  Among other cosmological models with
$w<-1$ and free of instabilities  that have been constructed we can mention the Lorentz-violating Dark Energy model \cite{Rubakov},
the ghost condensation model \cite{GostCond}, model with the kinetic gravitational brading \cite{vikmanmail}, model having the dust of the Dark Energy \cite{Lim:2010yk},
NEC violating models based on the Galileon theories \cite{Creminelli:2010ba}, and
the brane-world models \cite{Brane}.\footnote{A remark on stability is necessary. Ghost condensate models look unstable and this is cured by adding higher derivative terms. This in turn implies the presence of other ghosts and related  instabilities \cite{Emparan:2005gg,Kallosh:2007ad}. 
Nevertheless, since  a cutoff is assumed  and the wavelength of this ghost is smaller than the cutoff,  the  instabilities can be considered irrelevant.
The stability claim therefore relies on the fact that a well behaved UV-completion is assumed to exist. This is actually the case in \cite{Rubakov}, but it seems that  ghost condensate models cannot be UV-completed  in a usual way \cite{Adams:2006sv}.}

Other  results  from non-local cosmological models  have been obtained 
in the  context of inflation \cite{nongauss,Calcagni}.  Such models have in general the remarkable property that slow roll
inflation can proceed even in presence of an extremely steep potential \cite{Cline1}.  Furthermore they can 
produce non-gaussian signature for the cosmic microwave background. It has been shown in
\cite{nongauss}  that the non-linearity parameter which characterizes
this  non-gaussianity  can be
large,  as opposite to what is described by the canonical single scalar field inflationary models,  and
observationally distinguishable, for instance,  from Dirac-Born-Infeld inflation models.

Recently an attempt  to build  a non-local generalization of  Einstein gravity 
has also been developed in the literature \cite{Biswas_et_all} (see also \cite{Calcagni2} for an application of the
diffusion equation method to non-local gravity models).
One result obtained in this direction  is the construction of  an exact solution
which describes a non-singular bounce at the origin
and a  de~Sitter phase in the large time asymptotics.

The main focus of this paper is on  a particular class of cosmological models
that naturally arise in certain regimes of SFT \cite{IA1,nongauss,Calcagni,Cline1,Biswas_et_all,KVlast,Koshelev:2007fi}.  As it will be explained in details in section \ref{sftnl}, such models effectively
contain (infinitely) many scalar fields,  some of  which are characterized by the unusual feature of having complex masses squared.
Solutions to this type of models have been analyzed from the
cosmological point of view and have  proved to be interesting in order to address
the questions of the non-gaussianity and the phantom divide crossing.
It is however vital to consider the evolution of the (at least linear) perturbations. This step has not been yet covered in full
generality for inhomogeneous perturbations in the presence of fields with complex masses squared.
It  has been explored  only to some extent in  \cite{Cline1,KVlast,GKbog,KoshelevLast}, where space homogeneous
perturbations were analyzed.  The goal of this paper is to fill  this gap.

The paper is organized as follows.
In section \ref{sftnl} we review the general properties of this class of SFT derived models.
In section \ref{model} we present the specific model that we are going to consider  
and we show that it admits a class of simple exact solutions in the cosmological
 background represented by  spatially flat FRW universe. The general equations
describing the evolution of  cosmological perturbations around  these exact background  solutions are presented in section \ref{pert}.
In section \ref{two} we specialize to the case corresponding to a  truncation of the 
local system to a system with two complex conjugate fields with complex conjugate masses. In section \ref{numeric} we undertake  the numerical study of 
the perturbations and describe the results obtained. Section \ref{ghosts} describes the issue of  ghosts in our model.

\section{SFT based non-local models}
\label{sftnl}

Non-local models arise naturally in SFT \cite{review-sft} where,  under certain
approximations, one gets the following space-time action
\begin{equation}
S=\int dx \Big(\frac12\Phi\Fc_0(\Box)\Phi-V_{int}(\Phi) \Big) \, .
\label{mainaction}\end{equation}
Here $\Fc_0(z)$ is an analytic function on the complex plane with real
coefficients in the Taylor expansion. $\Phi$ is a scalar field from the
string spectrum and can be for instance the string tachyon.
To  be more specific consider for example cubic SFT, which may be either bosonic or fermionic.
Performing standard SFT  calculations one arrives to the
action for space-time fields. This consists of two parts: the quadratic
part, which is the kinetic term, and the cubic interaction among all the fields.
Schematically we can write it as
\begin{equation}
S=\int
dx\left(\frac12\phi_iK_{ij}(\Box)\phi_j-v_{ijk}(e^{-\frac{\beta}
2\Box}\phi_i)(e^{-\frac{\beta}2\Box}\phi_j)(e^{-\frac{\beta}2\Box}\phi_k)\right)\, ,
\label{action0}
\end{equation}
where  $\beta$ is a parameter
determined exclusively by the conformal field theory and $\Box$ is the
d'Alambertian operator. It is natural to
expect $\beta<0$, corresponding to  convergent Feynman graphs at large momenta.
From this specific example we see the non-local character of the action that is a general feature of SFT based models.\footnote{Appearance of higher derivatives is not an exclusive  feature  of this theory.
Non-commutative theories, for instance, also have higher derivatives, but these
non-local structures are very different.}
 If furthermore one integrates out all the modes of action  \eqref{action0} except one and if only the low mass
  excitations of  the SFT are retained, one obtains
\begin{equation}
S=\int
dx\left(\frac12\phi (\Box-m^2)\phi-\lambda\left(e^{-\frac{\beta}
2\Box}\phi\right)^p\right)
\label{action1}
\end{equation}
as an effective action. With a simple field redefinition $\Phi=e^{-\frac{\beta}
2\Box}\phi$ one obviously gets an action of the type outlined above.

The presence of the operator  $\Fc_0$ is the key modification with respect to a canonical scalar field action and its properties would play the most important role for the resulting model.  In SFT this operator can be computed perturbatively  by means of  the so-called ``level truncation scheme'' \cite{review-sft}. In general the action of the form \eqref{mainaction} is just an approximation. In principle a more involved entanglement of non-local form factors and scalar fields would be obtained. An example of  the operator $\Fc_0$ is given by what one obtains form the tachyon field in the low-level approximation, where one may have $\Fc_0=(\Box-m^2)e^{2\beta\Box}$. A specific expression for  $\Fc_0$  does not however simplify the following analysis, especially in the case of a curved background, and therefore we keep it general. Moreover, this operator may get modified if we couple our model with other ingredients.  Since this may result in possible restrictions on the operator and coupling structures in the theory, it is  important to keep $\Fc_{0}$  general.

From the SFT point of view,  the full potential $V=-\frac12\Fc_{0}(0)\Phi^2+V_{int}$   in \eqref{mainaction}  is expected to have at least two
distinct extrema. The scalar field exhibits a non-perturbative transition from one extremum to the other
and eventually stops.  The best understood processes of this kind in the SFT dynamics is the so called
rolling tachyon (see, for instance, \cite{AJK} and refs. therein), described as a time-dependent kink-like
solution. Several solutions in flat background were constructed numerically
\cite{AJK} and in some approximations analytically \cite{Vladimirov}. 
Taking the linearization of the potential around
a given extremum, say
$\Phi=\Phi_0+\varphi$, a significant analytic progress can be achieved.  The linearized  action
becomes quadratic in the scalar field $\varphi$ 
\begin{equation}
S=\int dx \Big(\frac12\varphi\Fc(\Box)\varphi \Big) \, .
\label{mainactionlocal}\end{equation}
In the linear approximation,  the action will actually have such a form and no other structures are expected.

For the action \eqref{mainactionlocal} one can establish an equivalence of
Lagrangians \cite{Koshelev:2007fi} (notice that this mapping is possible only for the quadratic linearized action  and not in presence of a potential, like in \eqref{mainaction})
\begin{equation*}
\varphi\Fc(\Box)\varphi \Leftrightarrow \sum_i\tau_i(\Box-J_i)\tau_i  \, ,
\end{equation*}
where in the r.h.s. the  $J_i$ are roots of the characteristic equation
$\Fc(J)=0$ and there are as many
terms in the series as roots. In other words there is a mapping of a non-local theory with one
non-local scalar field to a local theory with many local scalar fields $\tau_{i}$.
To each root is associated one local field, thus in the case of many but finite roots one obtains a genuine local theory with many scalar fields. When infinite roots are present, the non-locality of \eqref{mainactionlocal} is rephrased in the  appearance of an infinite number of local fields. In
general the roots  $J_i$ may be complex and, in this case, they appear in
complex conjugate pairs.
As a matter of fact this is not a problem for the theory, since the local  fields $\tau_{i}$ are non-physical. 
They only provide a different mathematical description. It is the original
non-local field  and the related quantities that must be real,
since they represent physical excitations. This translates in requiring  that the hermiticity 
of the local action is preserved when a pair of complex conjugate roots is present.  
This means that the fields $\tau_i$ also enter the
action in  complex conjugate pairs. The point we want to emphasize is that
instead of the usual $\tau\tau^*$ quadratic combination  we have terms of
the form
\begin{equation*}
\tau_i(\Box-J_i)\tau_i+\tau_i^*(\Box-J_i^*)\tau_i^* \, .
\end{equation*}
This quadratic expression in general  cannot be diagonalized in terms of real
fields and
this is one of the distinguishing features of such models compared to canonical
ones. 
It is however simpler to work with (infinitely) many local fields rather than one
non-local field.

The only thing which is important is  the particular vacuum around which the action is linearized and the associated spectrum of the theory.
There are many different possible vacua in string theory and the class of models we deal with practically distinguishes these vacua through  the form of the  operator $\Fc$. The latter is in turn characterized by its roots $J_i$ and there may be vacua in which we have no zeros at all  (it is believed that the true tachyon vacuum is of this kind with, for example, $\Fc=e^{2\beta\Box}$).


\section{Exactly solvable model}\label{model}

Looking for a cosmological scenario and expecting a similar behaviour in a slightly curved background, one is lead to consider  a scalar field that interpolates in between of distinct vacua of the potential in a curved background and eventually stops (i.e. $\varphi$
tends to $0$). Then, in the vicinity of the final vacuum,  one can construct approximate background solutions
for a model
where many scalar fields are minimally coupled to gravity and all $\tau_i$
tend to $0$ \cite{Koshelev:2007fi}. Furthermore, it was explicitly shown in \cite{GKbog} that exists a
simple generalization of such a model with (infinitely) many scalar fields
minimally coupled to gravity such that the equations of motion in the spatially
flat 
Friedmann-Robertson-Walker (FRW) metric become exactly solvable. This generalization
consists of an additional quartic potential
\begin{equation*}
\frac{3\pi G}2\left(\sum_i\alpha_i\tau_i^2\right)^2
\end{equation*}
that makes the full equations exactly solvable for
specific values of the parameters $\alpha_i$. (It was argued in \cite{GKbog} that
such a modification is obviously the next to
the quadratic order expansion of
the original potential around an extremum in the cases where the potential is
even around this
extremum, i.e. in cases where the expansion starts with
$V=\frac12m^2\varphi^2+\frac1{4!} \lambda \varphi^4+\dots$).
Such an additional term vanishes more
rapidly than the other terms of the original model  when evaluated on solutions
with the scalar fields going to zero. Therefore, in the far asymptotic
regime, the results for the model with and without the quartic extra term should coincide.\\

To be precise, the model of primary concern in the present paper is described by the
following action
{\small \begin{equation}
S = \int\!\!d^4x\sqrt{-g}\Biggl(\frac{ R}{16\pi
G_N}+\frac{1}{g_o^2}\Biggl(-\sum_i\frac{1}{2}\left(g^{\mu\nu}\pd_\mu\tau_i\pd_\nu\tau_i
+{J}_i\tau_i^2\right)-
\frac{3\pi G}2\left(\sum_{i}\alpha_i\tau_i^2\right)^2\Biggr)-\Lambda \Biggr)\, , 
\label{action_model_localexact}
\end{equation}}where  in addition to the quadratic piece a specific potential of the fourth degree is
present.  We work in $1+3$ dimensions with the signature
$(-,+,+,+)$, the coordinates are denoted by
Greek indexes $\mu,\nu,\dots$ running from 0 to 3. Spatial indexes are
$a,b,\dots$ and they run from 1 to 3. 
$G_N$ is the Newtonian constant, $8\pi G_N=1/M_P^2$, where
$M_P$ is the Planck mass, $g_o$ is the open string coupling constant, $G\equiv G_N/g_o^2$. $g_{\mu\nu}$ is the metric tensor, $R$
is the scalar curvature, $\Lambda$ is a constant.
The fields are dimensionless, while $[g_o]=\text{length}$.

We impose a specific relation between the parameters, namely $J_i=-\alpha_i(\alpha_i+3H_0)$, with  $H_0=\sqrt{\frac{8\pi  G_N\Lambda}3}$.  Such a relation, even not being expected to hold always, is the one that one gets in SFT considering the tachyon vacuum 
in the low-level approximation and with $dS$ background metric \cite{review-sft,Lyuda}.
Originally, because of the connection with SFT models, one would set the $J_{i}$ as the
fundamental family of parameters. On the other hand the modification
introduced here emphasizes the role of $\alpha_{i}$ as opposite to
$J_{i}$. Consider for example one specific $J_{i}$. In general there are two possible
$\alpha$ associated to it. If one wishes to include both of them in the action  one realizes that, due to the
specific form of the quartic potential, the two fields corresponding
to the two different  $\alpha$ are really independent. For
instance they cannot be introduced through their linear combination, as
opposite to the case of free fields with the same mass. They simply
need to be introduced in the action as two general fields, with the
special feature that the associated $J$ turns out to be the same. 
Moreover, as it will become clear, the exact solution is really
parametrized by $\alpha$. Finally, as it will be  discussed, for our purposes we are
specifically interested  in a particular  range of values for  $\alpha$. This practically associate in an
automatic way one single $\alpha_{i}$
to each  $J_{i}$ in most of the cases.

In the sequel we are going to explore the case of  spatially flat FRW cosmology, with our scalar fields system 
minimally coupled to the background metric
\begin{equation*}
g_{\mu\nu}=\diag(-1,a^2(t),a^2(t),a^2(t)) \, ,
\end{equation*}
where $a(t)$ is the scale factor, $t$ is the cosmic time and we shall use the
Hubble parameter $H(t)=\dot a(t)/a(t)$, denoting with the dot the derivatives
w.r.t. the cosmic time. In such a background the modified local action has the following equations of motion
\begin{equation}
\label{FrEOMgFRWex}
\begin{split}
3H^2&=4\pi G\left(\sum_i\left(\dot\tau_i^2
+{J}_i\tau_i^2\right)+3\pi G\left(\sum_{i}\alpha_i\tau_i^2\right)^2\right)+8\pi G_N\Lambda\, ,\\
\dot H&=-4\pi G
\sum_i\dot\tau_i^2 \,  ,\\
\end{split}
\end{equation}
and
\begin{equation}
\ddot\tau_i+3H\dot\tau_i+J_i\tau_i+{6\pi G}\alpha_i\tau_i\sum_{j}\alpha_j\tau_j^2=0\, ,\qquad \text{for all}~i  \, .
\label{FrEOMtauFRWex}
\end{equation}
It is easy to check that they admit the following  exact solution
\begin{equation}
\begin{split}
\tau_i&=\tau_{i0}e^{\alpha_i t}\,  ,\\
H&=H_0-2\pi G\sum_i\alpha_i\tau_{i0}^2e^{2\alpha_i t} \, .
\end{split}\label{action_model_localexactsol}
\end{equation}
This solution is valid for any number of fields (including a single field and infinitely many fields)
and for any values of the parameters $\alpha_i$ (i.e. real or complex).
We see that if $\text{Re}(\alpha _i)<0$ for all $i$ then solutions vanish and moreover the quartic
term in the potential practically vanishes at sufficiently shorter times and we
are left with free fields.
Thus for large times the model without a quartic
potential is restored and we return to more general SFT based models without
requiring any other conditions on
the potential to be satisfied.
 More generally, \eqref{action_model_localexactsol} is a solution of linearized Friedmann equations
in any non-singular potential with $\tau_i=0,~H=H_0$ as a stable fixed point, according to Lyapunov theorem, if $Re(\alpha_i)<0$. In our particular setting we deal with a potential for which this solution is exact for the full action. This would simplify the task of considering perturbations, but a generalization of the results to other forms of the interaction seems  straightforward.

From the physical point of view, the solutions above describe a  very interesting phenomenon,
the phantom divide crossing. If there are complex roots $J_{i}$, $H$ oscillates
around the values $H_{0}$
and the equation of state parameter $w = p / \rho$ crosses the value $w=-1$ ($p$ and $\rho$ are the 
 collective  pressure and energy density). Indeed 
\begin{equation}
 w=-1-\frac{2}{3}\frac{\dot H}{H^{2}}\, ,
\end{equation}
and $\dot H$ is an alternating function.
Notice that such a crossing is generically not possible with single ordinary
scalar field.
With two scalar fields one can obtain such a behavior, but one of the two fields must be a phantom (ghost) in this case. 
In our setup, in presence of complex roots, we also have at least two fields,  
which are however just an effective description of the original theory with a
single non-local field. Moreover, recall that the complex fields enter the
quadratic part of the Lagrangian in 
the following way:
\begin{equation*}
\tau_i(\Box-J_i)\tau_i+\tau_i^*(\Box-J_i^*)\tau_i^* \, .
\end{equation*}
This can be recast in terms of real components as
\begin{equation}
\chi_i \Box
\chi_i-\psi_i\Box\psi_i-m_i(\chi_i^2-\psi_i^2)+2n_i\chi_i\psi_i \, ,
\label{toy}
\end{equation}
where $\tau_i=\frac{1}{\sqrt{2}}(\chi_i+i\psi_i)$ and $J_i=m_i+in_i$.
This Lagrangian would be a partial case of the so called B-inflation model \cite{AnBaVi}
provided the ``mass'' $m_i$ becomes zero for one of the fields, say $\psi_i$.
Also, this resembles the so-called quintom model \cite{Quinmodrev1} where two fields (one normal and
one phantom) are used as well, but without a quadratic coupling.
Because of this coupling we cannot exactly identify  the states as
normal or tachyons or ghosts, since these notions refer to diagonalized Lagrangians with real fields
(in our case we cannot make it diagonal with real fields).
This does not mean however that the issue of ghosts is not present in our model. We will return to this question in section \ref{ghosts}.


\section{Cosmological perturbations}\label{pert}

We want to consider the problem of cosmological perturbations for the 
exact solutions  presented in the previous section. We focus on the
scalar type perturbations. These are induced by energy density
inhomogeneities and are the one that may cause cosmological
instability, as opposite to vector type  perturbations, that are known to
decay rapidly and tensor type perturbations, that correspond to
gravitational-wave perturbations of the metric and do not induce any
perturbation in the perfect fluid at the linear order. The three kind
of  perturbations are decoupled and thus one can safely consider them separately \cite{Mukhanov}.

We start presenting the equations for spatially inhomogeneous
perturbations of the modified action introduced in the previous
section. We refer the reader  to \cite{Mukhanov,bardeen,hwangnoh} for the
details on the general derivation  and  to \cite{KVlast}  for the conventions
adopted here.
In our setup the primary quantities are the scalar fields $\tau_i$. It is
therefore more transparent to use the fields themselves rather than the notions of
energy and pressure for the corresponding components of the stress-energy
tensor.
It is also important that the spatial dependence of the perturbation variables
can be represented as $e^{ik_ax^a}$ for the spatially flat FRW universe,
where ${k_a}$ is a 3-dimensional vector and $k^2=\delta^{ab}k_ak_b$. With this
representation all the relevant information about the space-inhomogeneities are
collected into the
comoving wavenumber $k$.

After a number of tedious manipulations
the relevant equations for 
scalar perturbations read \cite{KVlast,hwangnoh}
\begin{equation*}
\begin{split}
\ddot\zeta_{ij} &+ \dot\zeta_{ij}\left( 3 H +
  \frac{\ddot\tau_{i}}{\dot\tau_{i}} +
  \frac{\ddot\tau_{j}}{\dot\tau_{j}} \right)  + \zeta_{ij} \left(
  -3 \dot H + \frac{k^2}{a^2} \right) =  \left[ \frac{\tau_{i}}{\dot\tau_{i}} \left( J_{i}  + 6 \pi G \alpha_{i}
    \sum_{l} \alpha_{l} \tau^{2}_{l} \right) \right. \\
 & \left. -  \frac{\tau_{j}}{\dot\tau_{j}} \left( J_{j}   + 6 \pi G \alpha_{j}
    \sum_{l} \alpha_{l} \tau^{2}_{l} \right)   \right] \left( \sum_{k}
  \frac{\dot \tau^{2}_{k}}{ \rho + p }\left( \dot\zeta_{ik} +
   \dot\zeta_{jk}\right) + \frac{2 \varepsilon}{1+w}\right) \\ 
 &+ 12 \pi G \sum_{k} \alpha_{k} \tau_{k} \dot\tau_{k} \left( \alpha_{i}
  \frac{\tau_{i}}{\dot\tau_{i}} \zeta_{ik}  -  \alpha_{j}
  \frac{\tau_{j}}{\dot\tau_{j}} \zeta_{jk}\right) 
\, ,
\end{split}
\end{equation*}
\begin{equation*}
\begin{split} 
&\ddot\varepsilon + \dot\varepsilon H \left( 2 - 6 w + 3 c^{2}_{s}
\right) + \varepsilon \left( \dot H (1 - 3 w) - 15 w H^2 + 9 H^2
  c^{2}_{s} + \frac{k^2}{a^2} \right)  \\
&\quad = \frac{k^2}{a^2} \frac{1}{\rho} \frac{2}{\rho + p } \sum_{k,l}
\left(J _{k }  + 6 \pi G\alpha_{k}\sum_{j} \alpha_{j} \tau^{2}_{j}\right)
\tau_{k}\dot\tau_{k} \dot\tau^{2}_{l} \zeta_{kl} \, ,
\end{split}
\end{equation*}
where $\zeta_{ij}=\frac{\delta\tau_i}{\dot\tau_i}-\frac{\delta\tau_j}{\dot\tau_j}$ is
the gauge invariant variable for the scalar fields perturbations and  $\varepsilon$ is the
gauge invariant total energy density perturbation.  In writing the equations we
have  introduced the definition   $c_{s}^{2} = \dot p / \dot\rho$ for the
speed of sound. 

Equipped with these general equations we can consider as a background the exact class
of solutions \eqref{action_model_localexactsol}.  Substituting the
expression for $\tau_i$, one gets some cancellations and
simplifications. Keeping track of all of these, the final result is:  
\begin{equation}
\begin{split} 
\label{perturbexactzeta}
\ddot\zeta_{ij} &+ \dot\zeta_{ij}\left( 3 H +
  \alpha_{i} +
  \alpha_{j} \right)  + \zeta_{ij} \frac{k^2}{a^2}   =    \frac{\left(
  \alpha_{i}  -  \alpha_{j}  \right)}{\dot H}  \\
  &  \left( 4 \pi G \sum_{k}
 \tau^{2}_{0k} \alpha^{2}_{k} e^{2 \alpha_{k}t}\left( \dot\zeta_{ik} +
    \dot\zeta_{jk}\right) + 3 H^2 \varepsilon\right)\, ,  
\end{split}
\end{equation}
\begin{equation}
\begin{split} 
 \label{perturbexacteps}
\ddot\varepsilon & + \dot\varepsilon \left( 5 H  + 4 \frac{\dot H}{H}
  - \frac{\ddot H}{\dot H} \right) + \varepsilon \left( 6 H^2  + 14
  \dot H + 2 \frac{\dot H^2 }{H^2}   - 3 H  \frac{ \ddot H}{\dot H}+
  \frac{k^2}{a^2} \right)     \\
&= \frac{k^2}{a^2} \frac{4}{3} \frac{(4 \pi G)^2}{\dot H H^2 } \sum_{k,l}
  \alpha_{k} \alpha^{2}_{k} \alpha^{2}_{l} \tau^{2}_{0k}\tau^{2}_{0l}
 e^{2\alpha_{k}t} e^{2\alpha_{l}t} \zeta_{kl}\, ,  
 \end{split}
\end{equation}
where the equations of  state parameter and the speed of sound have been
expressed in terms of the Hubble parameter and its derivatives,
according to Friedmann equations. 

A general analysis of the system  given by \eqref{perturbexactzeta} and \eqref{perturbexacteps}
is difficult, even in the case where the background solutions  have the
simple form \eqref{action_model_localexactsol}.
Many models with multi scalar fields derived
from non-local model with a single scalar field  and the
related problem of cosmological perturbations  have already 
been considered in the literature. Nevertheless, for all the motivations discussed
above, the model presented here represent to a certain extent a
novelty in this context. In the analysis of the cosmological
perturbation we will therefore concentrate  on the simplest case that
capture the interesting features of the model presented in this
work, namely the case of two complex scalar fields. 

 \section{Two complex conjugate roots}\label{two}

When considering complex solutions one should remember  that the physical
quantities that appear in the original model must be
real. In the case of two complex conjugate solutions, labelled by the parameters
$\alpha_{1}$ and $\alpha_{2}=\alpha_{1}^{*}$,
that we wish to consider, this amount to choose the integration
constants in such a way that $\tau_{1} = \tau_{2}^{*}$. This is exactly the setting we are going to consider.
Using the representation  $\alpha_{1} =\alpha_{2}^{*} = \alpha = -\frac{x}{2} + i
\frac{y}{2}$ or $\alpha=|\alpha|e^{i\phi_\alpha}$ and using the explicit solutions for $\tau_{1}
=\tau_{2}^{*}  =\tau_0e^{\alpha t} $ one can write all the quantities appearing in the
differential equations by means of trigonometric functions.
 Upon substitution and after some algebra
 the equations of interest take the form:
\begin{align}
\ddot\chi +  \dot\chi  \left( 3 H + x + y \tan( y t) \right) + \chi
\left( 3 x H + y x \tan(y t) +\frac{k^2}{a^2} \right)
 = \frac{ y \varepsilon H^{2}}{\cos(yt)} \, , \label{pert_chi}
\end{align}
\begin{align}
\ddot\varepsilon &+  \dot\varepsilon  \left(  5 H  + 4\frac{\dot
      H}{H}  +x + y \tan (y t  )\right) \nonumber \\ 
      &+ \varepsilon \left( 6 H^2  + 14 \dot H +  2\frac{\dot
      H^{2}}{H^{2}} + 3x H     + 3 y H\tan( yt ) +\frac{k^2}{a^2} \right)
 = -\frac{k^2}{H^2 a^2}  \frac{y \chi}{\cos(yt)} \, . \label{pert_eps}
\end{align}
Note that all the quantities appearing in the equations are real. All
the arbitrary phases have been eliminated. This can be done in
all generality corresponding just to a shift in the time
variable.
The function $\chi$ is also real and is related to the
original perturbation parameter $\zeta_{12}$ for the scalar fields as
$\chi =  i \frac{8 \pi G |\alpha^2| |\tau_{0}|^{2}
}{3}e^{-xt}\zeta_{12}$.
Notice that one can always take $y>0$ without any loss of generality.
In fact one sees from the explicit form of the equations that a change of sing for $y$ can be always reabsorbed
as an extra minus factor in the definition of  $\chi$. This is  not important if we are interested  only in
separating regions in the parameters space where the perturbations are confined from regions where they are not.
Moreover, $y$ just sets the periodicity of the functions appearing in the differential equations
and can be therefore conveniently rescaled when studying the numerical evolution of the perturbed system.
This   corresponds to a change of variable from the cosmic time  $t$ to the dimensionless time coordinate $T = y t $. 
With such a rescaling of the time variable the perturbative equations are characterized 
by a set of dimensionless parameters,   $H_{0}/y$, $\tau_{0}$, $x/y$ and $k/y$,
once an extra $1/y$ factor is reabsorbed in the definition of $\chi$.  Note that with this  redefinition of the time variable
one also has $H(t) = yH(T)$.
From a practical point of view the rescaling  is completely equivalent to set $y=1$ in \eqref{pert_chi} and \eqref{pert_eps}, with 
a set of parameters, $H_{0}$, $\tau_{0}$, $x$, $k$, that are now dimensionless. 
We therefore adopt this choice,  setting $y=1$ hereafter and we still denote the time variable as $t$.
The explicit expression for $H$  in the parameterization adopted is: 
\begin{equation}
H = H_0+h=H_{0} - 4\pi G |\alpha||\tau_{0}|^{2} e^{-x t}
\cos(t - \phi_{\alpha}). \label{Htrigonometric}
\end{equation}
This makes transparent that for certain parameters one can have 
an expanding universe regime, $H>0$, with $H$ reaching the constant
value $H_{0}$ at large times.
It is also useful to keep in mind that 
\begin{equation*}
\dot H = - 8\pi G |\alpha|^2|\tau_{0}|^{2} e^{-x t}
\cos( t) 
\end{equation*}
and
\begin{equation*}
a = a_o\exp\left(H_0t- 2\pi G |\tau_{0}|^{2} e^{-x t}
\cos( t-2\phi_\alpha)\right) \, .
\end{equation*}
We stress one more time that hereafter all the quantities appearing in the equations and in the definition of $H$ and $a$
must be understood as rescaled dimensionless quantities.
Note also that we emphasized that we are particularly interested in situations 
where the quartic potential vanishes for large times, corresponding to  $x>0$.
As discussed, the rescaling does not arm this point and we can safely work in terms of positive
rescaled dimensionless $x$. 

The system of equations \eqref{pert_chi}-\eqref{pert_eps} is in
principle ready to be analyzed. An analytic solution for this problem
is however  not easy to find. In order to proceed with a numerical analysis it is worth noting
that  there are some special points in the evolution of the system
under consideration. Namely, when  $  t = \pi/ 2 + n \pi$  one has divergent coefficients appearing in the
differential equations. Indeed at these points $\cos(t)$
becomes zero. The numerical integration process may encounter some
problem when it reaches these points if the precision is not high enough. Even worse, the system could
be really singular. Expanding in series around such  points the differential
equations read:
\begin{equation} \label{taylorexp}
\begin{split}
&t \ddot \chi -   \dot\chi  - x \chi 
 =  - (-1)^{n}  H^{2}_{n} \varepsilon  \,  , \\
&t \ddot \varepsilon - \dot \varepsilon - 3  H_{n} \varepsilon =
(-)^{n}\frac{k^2}{ H^{2}_{n} a^{2}_{n}}\chi \,  , 
\end{split}
\end{equation}
where the constants  $ H_{n}$ and $ a_{n} $ are respectively the functions $H(t)$ and
$a(t)$ evaluated at the $n$-th singular point.  From \eqref{taylorexp}
it is quite clear that nothing happens close to the singularity. Numerical
integration of this equations and of the non-approximated ones close
to the singularity shows that one can pass the singular points without
any problem and, if the precision is high enough, all the features of
the differential system  are captured.


\section{Numerical study}\label{numeric}

We describe here the numerical results obtained starting
form the system of equations presented in the previous section
and the picture that emerges for the evolution of the perturbations 
when the different quantities characterizing the  background
and the perturbations themselves are varied.
We particularly focus on distinguishing between situations where 
the system remains stable and situations where 
the perturbations grow making the system unstable.
For simplicity we set the constants $G =1 / 8\pi$ and $a_{0}=1$.
This choice does not affect the evolution of the system in any way.

\begin{figure}[htb]
\begin{center}
$\begin{array}{cc}
\includegraphics[width=0.45\textwidth]{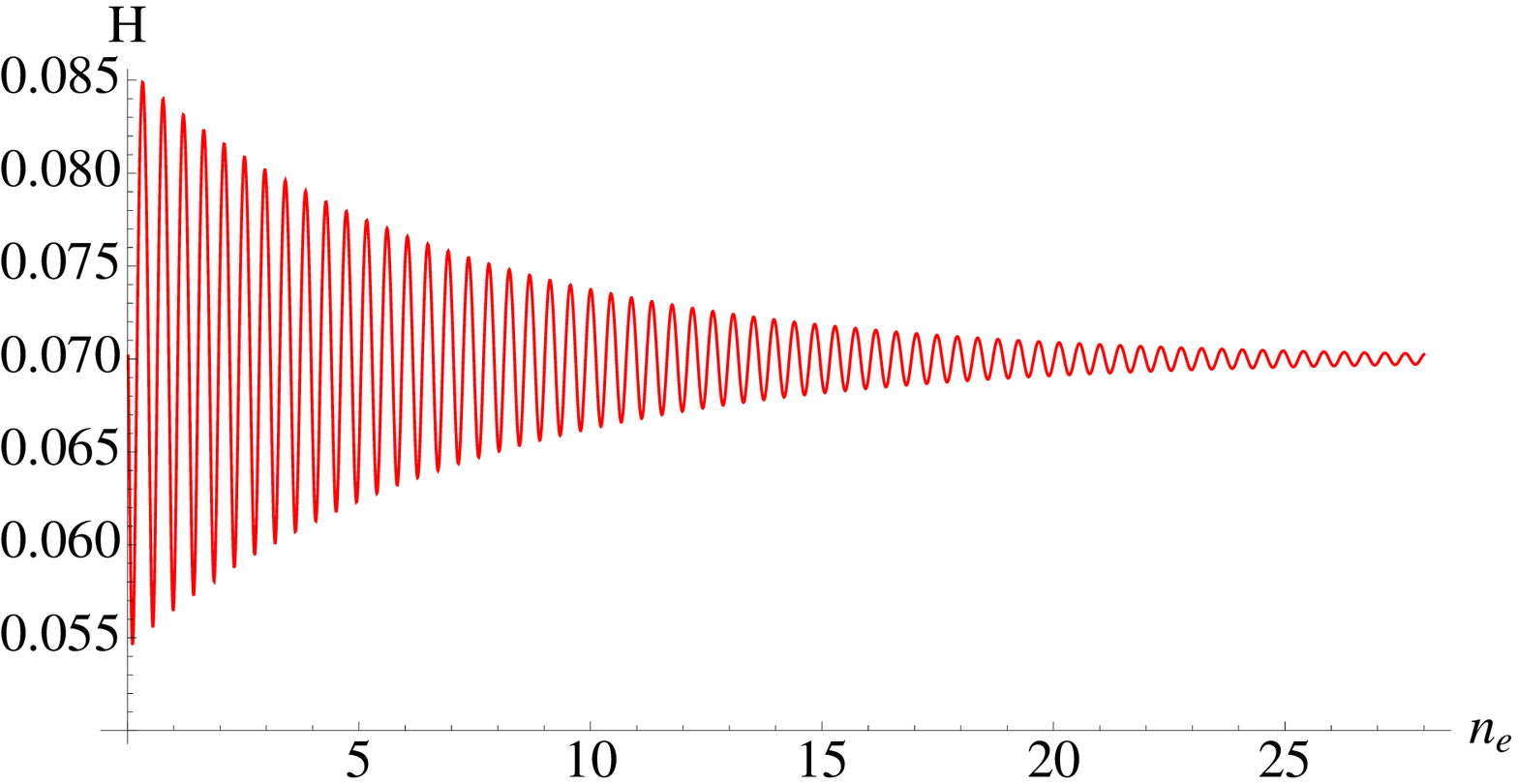} &
\includegraphics[width=0.45\textwidth]{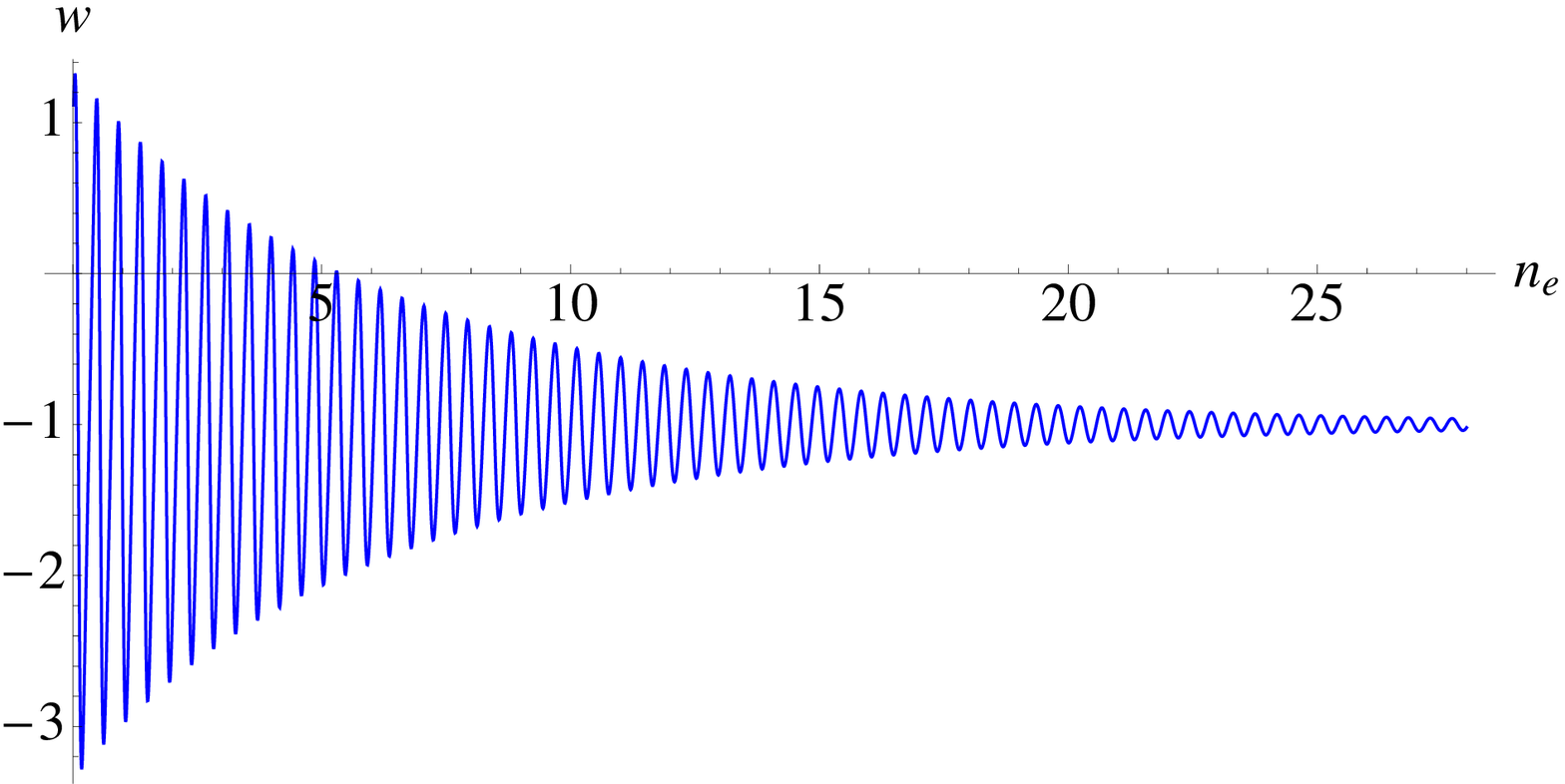}   
\end{array}$
\end{center}
\caption{In the left graph, the typical evolution of the Hubble parameter considered
  in our setting is plotted against the e-folding number. It evolves with damped oscillations towards the
  asymptotic value $H_{0}$,  never crossing the zero. In the right graph, the
corresponding evolution for the equation of state parameter $w$ is shown. The crossing of
$w=-1$ 
  that characterizes the case of complex roots is evident.  The value of the constant parameters considered in the plots are: $H_{0}=0.07, ~x=0.01,~\tau_{0}=0.25$. }
\label{fig:Hubble} 
\end{figure}

In order to numerically integrate the system of differential equations \eqref{pert_chi}-\eqref{pert_eps}
exact background solutions must be specified. In the case of two complex conjugate roots 
considered here, this corresponds to specify three real dimensionless parameters, $H_{0}$, $\tau_{0}$, $x$.
We specifically focus on the cases of asymptotically vanishing quartic potential, corresponding to positive $x$ parameters.
 Within this class of solutions, the Hubble parameter $H$ evolves with damped oscillations
around the constant value $H_{0}$. In the analysis of the perturbed system we concentrate on the case of cosmologies with 
expanding universe. This corresponds to  ask  $H >
0$  in all the stages of the evolution.
Such a request, in the region where $\text{Re}(\alpha)<0$, is easily fulfilled
with a balanced choice of $\alpha$ and $\tau_{0}$  with respect to $H_{0}$. In figure \ref{fig:Hubble} is shown the typical evolution of  the Hubble parameter considered in our study of cosmological perturbations.

One comment is necessary in this regard. The constant of integration $\tau_{0}$ can be in principle tuned at will for the purpose 
of reproducing the desired behaviour for $H$.
However, from a physical point of view, this choice is not completely arbitrary.
Equation \eqref{FrEOMgFRWex}  shows indeed that the value of $\tau_{0}$  is an important 
parameter in determining the origin of the Dark Energy content of the model.

One additional parameter that  characterizes the perturbed system  
and that must be specified in order to numerically study the evolution
of the perturbations is the  comoving wavenumber $k$.
This always enters  the differential equations in a specific way.
Namely, it always appears in the combination $k^{2}/a^{2}$
and it is therefore mainly relevant in the early stages of the
evolution. At large times it become less and less important, due to the 
suppression given by the increasing scale factor. 
In particular this combination enters the non-homogeneous term of the differential equations
and is therefore an important parameter in setting the coupling between 
the scalar fields perturbation parameter and the energy
density one. Note however that it appears in a non-symmetric way in the differential system \eqref{pert_chi}-\eqref{pert_eps}.
It only appears in the non-homogeneous term of $\eqref{pert_chi}$, while it enters in a similar way 
into the  left hand sides of the two differential equations.

Finally a set of initial conditions for the  differential problem 
must be specified in order to numerically integrate the differential equations.
In particular one would like turn on a small initial non-zero value
for one or both the quantities $\varepsilon$ and $\chi$,  at some given time.
The initial value problem requires also a set of initial conditions for the first time derivative
of the two perturbative parameters.

\begin{figure}[htb]
\begin{center}
$\begin{array}{cc}
\includegraphics[width=0.45\textwidth]{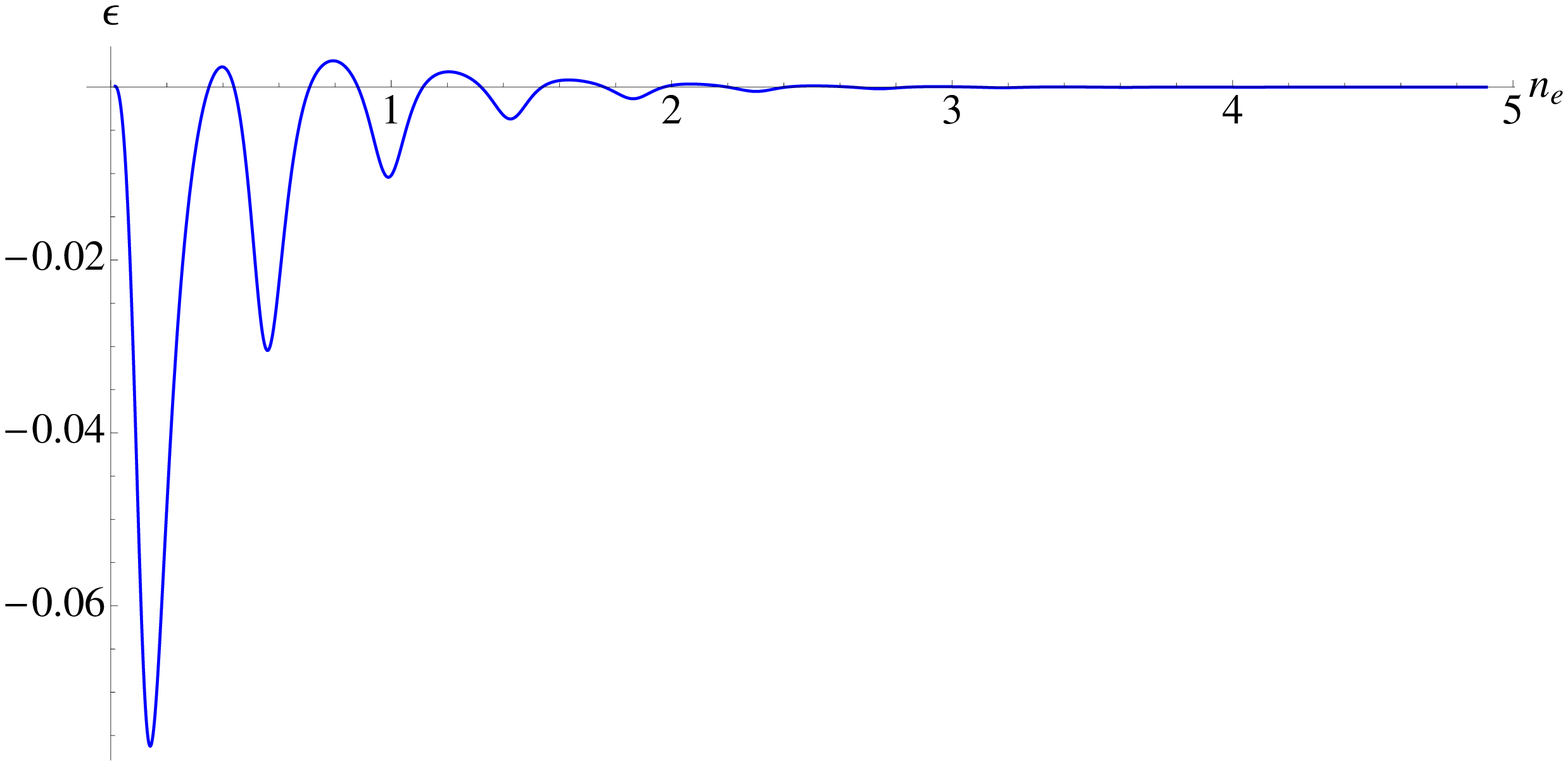} &
\includegraphics[width=0.45\textwidth]{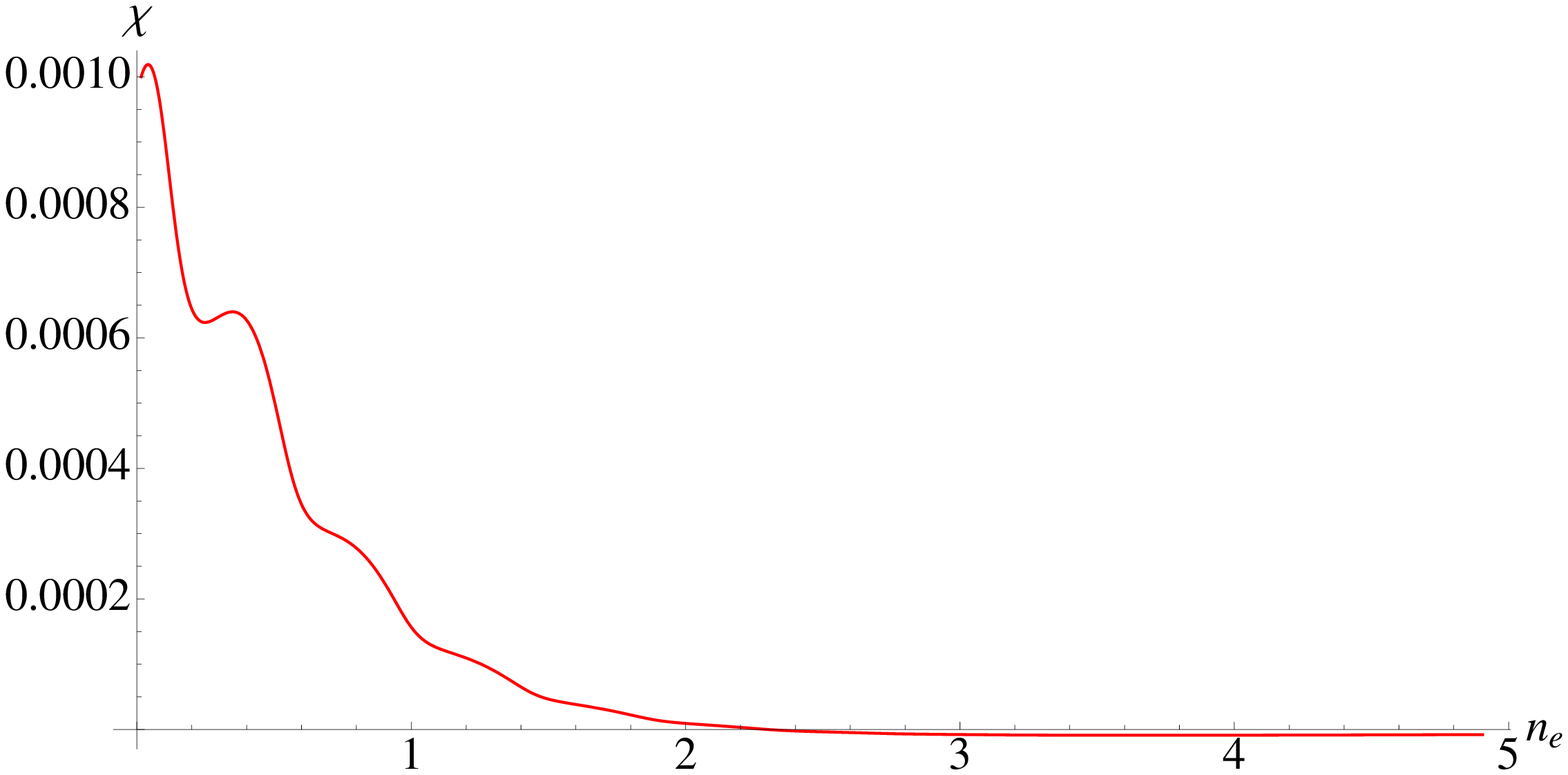}   
\end{array}$
\end{center}
\caption{ The evolution of $\varepsilon$ and $\chi$ for  $H_{0}=  0.07$, $x=
0.01$, $\tau_{0} = 0.25$, $k= 0.5$ is shown. The initial conditions are:
$\varepsilon(0) = 10^{-4}$, $\dot\varepsilon(0)=0$,  $\chi(0)= 10^{-3}$ and $\dot\chi(0)=10^{-4}$. }
\label{fig:epsnormal} 
\end{figure}

There are mainly two different situations emerging from the 
study of the  differential system \eqref{pert_chi}-\eqref{pert_eps}.
In both cases one observes that the perturbations decrease to zero 
at sufficiently large times, giving in principle a well behaved 
evolution for the system, such as the one shown in figure \ref{fig:epsnormal}. There are however particular regimes 
where one must pay attention. 
While in general one could say that the perturbations evolve
with  oscillatory behaviour according to various possible different paths,
but eventually always decreasing to zero value for both $\varepsilon$ and $\chi$,
in some particular cases one observes a fast initial growth for $\varepsilon$.
Precisely, one does not observe divergence for $\varepsilon$.
On the contrary, it eventually vanishes, 
but in the initial stages of the numerical evolution it can
exit the perturbative regime. 

\begin{figure}[htb]
\begin{center}
$\begin{array}{cc}
\includegraphics[width=0.45\textwidth]{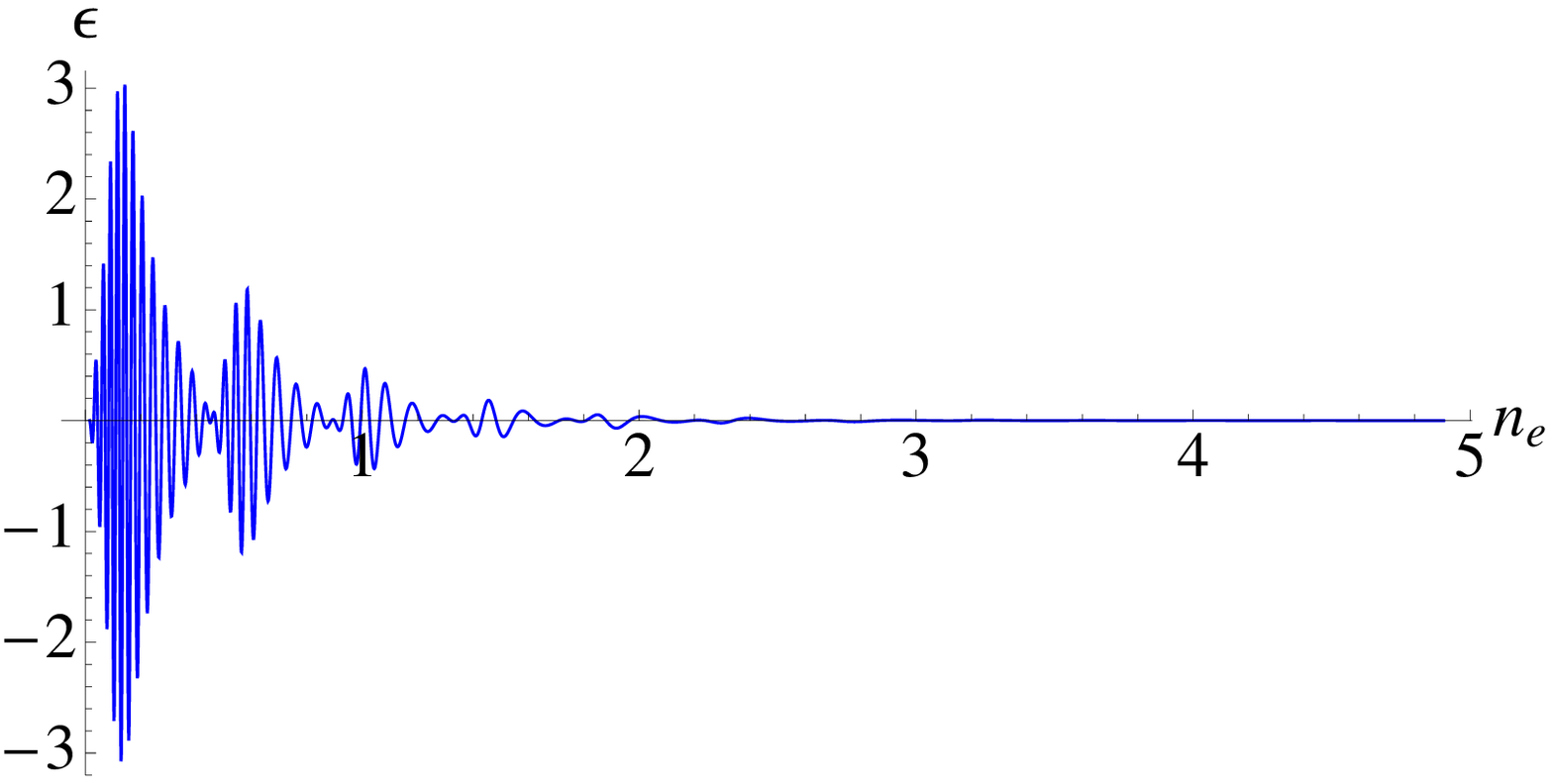} &
\includegraphics[width=0.45\textwidth]{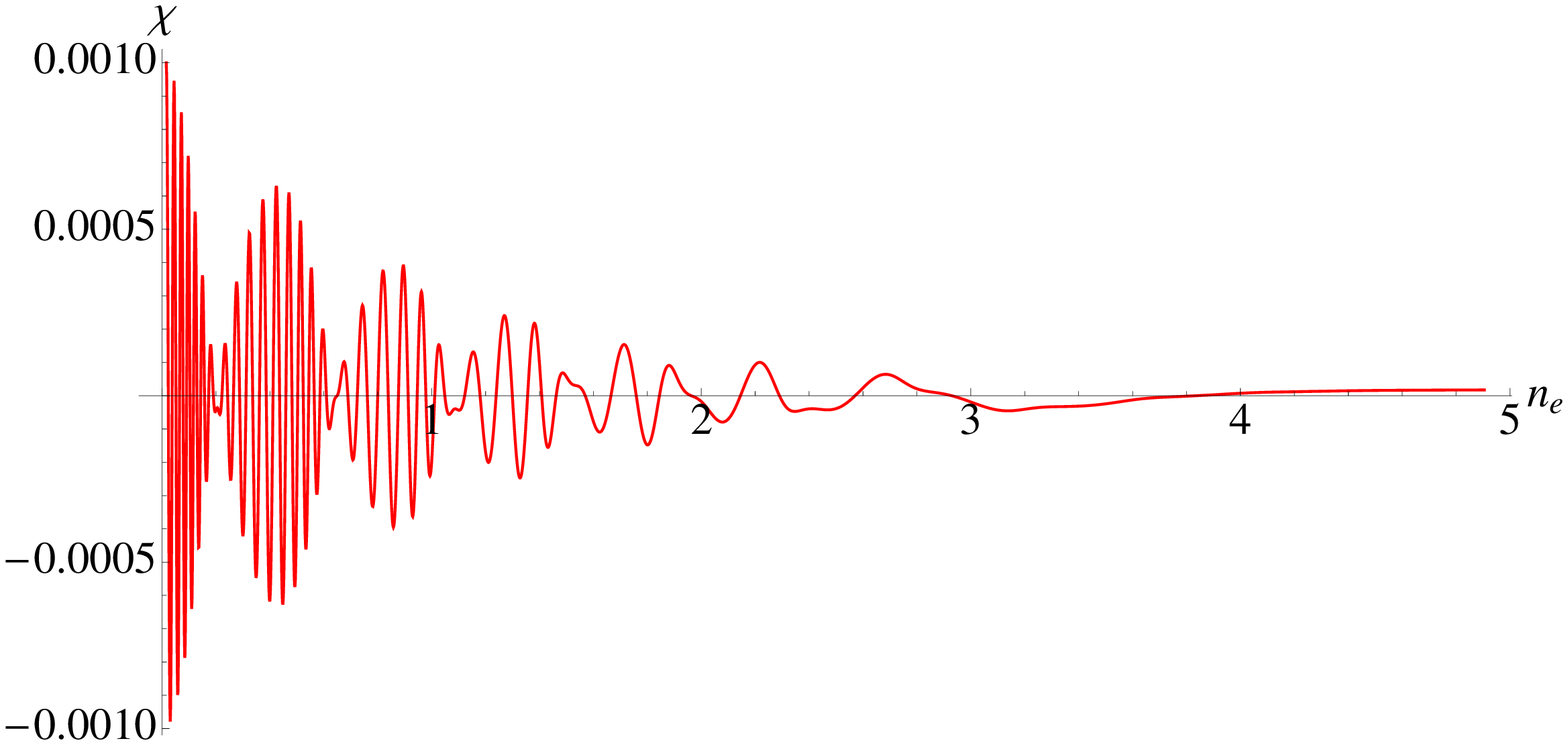}   
\end{array}$
\end{center}
\caption{ The evolution of $\varepsilon$ and $\chi$ for  $H_{0}=  0.07$, $x=
0.01$, $\tau_{0} = 0.25$, $k= 15$ is shown. The initial conditions are:
$\varepsilon(0) = 10^{-4}$, $\dot\varepsilon(0)=0$,  $\chi(0)= 10^{-3}$ and $\dot\chi(0)=10^{-4}$. Even though both $\chi$ and $\varepsilon$ stay confined and eventually decrease to zero value, $\varepsilon$ 
goes trough an initial phase where it shows a fast growth and exits the range of validity of the perturbative regime.}
\label{fig:epsbig} 
\end{figure}

Indeed when one has non-zero initial conditions for $\chi$, 
irrespectively of the initial value of $\varepsilon$ and its derivative,  one naively observes that 
if $H_{0} < 1$  and if $k^{2}/H_{0}^{2}$ is of the same order of magnitude 
of $1/ \chi$, for values of $\chi$ taken in the initial stages of the evolution,  $\varepsilon$  increases. 
If the non-homogeneous term in \eqref{pert_eps} is large enough,
the energy density perturbation parameter will exceed 
the range of values where the perturbative regime makes sense.
An example of such a situation is depicted in figure \ref{fig:epsbig}.
This means that if we start with non-zero initial conditions, one can always find a value of $k$ large enough
to trigger a fast and large growth for $\varepsilon$. In fact  the conditions naively identified for such a phenomenon to occur do just mean that the
non-homogeneous term should remain significantly  large in the initial stages of the evolution.
The scale factor appearing in the non-homogeneous term, on the other hand gives an exponential suppression of the latter  with a rapidity roughly set by $H_{0}$.
This  can be always compensated with an appropriately large value of  $k$, so as  for the balance of the other factors appearing in the non-homogeneous term of \eqref{pert_eps}.
The exponential suppression of the scalar field solutions, due to the positive values of $x$ considered here, also play a role in this subtle  
interplay of $k$ with the other parameters. In fact the amplitude of the oscillations covered by $\varepsilon$ increases roughly in a polynomial way with $k/x$. 
Specifically, for small values of $k/x$ it increases like $(k/x)^2$. When the ratio $k/x$ gets larger this power law behaviour is smoothed out and the maximum range of values spanned by 
$\varepsilon$ increase more slowly with $k/x$.

In the opposite case, where one sets  zero initial conditions  for $\chi$ and $\dot\chi$,
the spectrum of perturbations goes to zero without any particular surprise.
$\varepsilon$ evolves with different oscillatory modes, 
spanning a range of possible values  of the same order of magnitude of the initial conditions and eventually vanishes.
In these cases  $\chi$ is also excited, but  only up to 
values that are practically negligible. Such an asymmetric behaviour dictated by different initial conditions
shows that exists a strong transfer  from $\chi$
to $\varepsilon$, but not in the opposite direction.

One can  therefore conclude  that for any value of the parameters $H_{0}$, $\tau_{0}$, $x$ and for any  
initial conditions, provided that $\chi$ and $\dot\chi$ have not simultaneously zero initial value, 
it is always possible to find a $k$ large enough to obtain a rapid and large growth of $\varepsilon$.
This  can bring $\varepsilon$  out of the range of validity of the linear approximation adopted in the study of the cosmological perturbations.
When this happens one cannot relay anymore on the perturbative computations.
Since, however, the perturbations remain confined and eventually they all go to zero, it  is an interesting question 
whether this phenomenon signals a real instability of the system or if a more careful analysis,
such as considering higher order perturbations, could reveal  that the system is indeed stable irrespectively of the value of $k$. 

As it was explained at the end of section \ref{model}, the case of (two) complex conjugate
roots corresponds to a situation where the phantom divide crossing is realized.
The analysis of this section shows that for a large range of parameters the
system has a stable behavior under inhomogeneous perturbations
and the perturbations are in any case confined and vanish. This allows us
to claim that the exact solutions presented in this paper give a stable (at least up to linear perturbations)
model with the equation of state parameter
oscillating around the value $w =-1$.

\section{Taming the  ghosts}
\label{ghosts}

The numerical analysis of the previous section shows that at the linear order  in perturbations 
the system is stable  for a wide range of parameters and initial conditions. Even in cases where the perturbations grow, posing a question on the validity
of the perturbative approach, it is enough to wait a sufficiently long time to see that perturbations finally go to zero.  One could therefore conclude that there are no divergent modes in perturbations. This however  does not mean that ghosts, which are effectively present
 in our setup, would not trigger instabilities at higher orders in perturbations or when coupled to other fields \cite{Carroll:2003st,Cline:2003gs,Woodard:2006nt}. It is therefore  important to understand  to what extent  ghosts do not pose a threat for the system and our results are applicable.
 
To clarify the picture we return to the SFT motivation of the model we analyzed. We  have considered  the next-to-the-linear approximation of SFT dynamics around a given vacuum. There are many different possible vacua in string theory and the class of models we deal with practically distinguishes these vacua through  the form of the  operator $\Fc$. The latter is in turn characterized by its roots $J_i$. There may be vacua in which we have no zeros at all  (it is believed that the true tachyon vacuum is of this kind for example). Such a vacuum is definitely ghost-free. There may also be  vacua with only one root and in this case one has only field, canonical or ghost. When, however, more than one root is present ghosts appear in the spectrum.
There are therefore  two issues which should be clarified: what is the lifetime of such a vacuum before it gets destroyed by the presence of ghost and whether one can UV complete the model.

As outlined above,
 on the one hand one might expect that considering higher order perturbations would 
make manifest ghosts instabilities that are not seen at the linear level. On the other hand, looking at $\eqref{toy}$, it is clear that  the ghosts degree of freedom in our model 
is coupled not only through gravity with everything else, but also directly with $\chi$.
The instability is due to the infinite phase space associated to the arbitrarily large negative energies of the ghost that  result in a divergent decay rate for the vacuum. However, this instability might be not dramatic in an absolute sense. 
In fact,  the phantom theory is meant to be just an effective description of certain regimes here. The validity of this description 
therefore depends on whether the ghost instability has the time to develop or not over the  relevant cosmological time scale.\footnote{See for instance 
\cite{Cai:2007zv} for an example where the phantom phase is just an intermediate phase in the cosmological evolution. Some unstable modes propagate out of this phase and some other not.}  

We follow the analysis of  \cite{Carroll:2003st}  in order to give an estimate of the domain of validity of our specific model as an effective description with cutoff  $\Lambda_{cutoff}$. In \cite{Carroll:2003st} the allowed decay process  involving  phantom particles have been analyzed.  We are particularly interested to the situation where there is just one ghost field, so as we are assuming in our case.  Ordinary particles  can decay into ghosts plus other ordinary particles, provided that the effective mass of the final ordinary particles is larger than the original one. When  just one ghost is present,  this can decay only if in the final states there are at least two ghosts and, at least, an ordinary particle.
In order to compute the decay rate of our vacuum it is useful to rewrite the action involving the ghost field with the canonical normalization. We therefore perform the rescaling $\tau/g_{o} \rightarrow \tau$ in the action \eqref{action_model_localexact} where, in the case under consideration, the index $i$ denoting the different field run over one complex field $\tau$ together with its complex conjugate. We  then expand the scalar field Lagrangian in terms of the rescaled real field $\chi$ and $\psi$, with  $\tau=\frac{1}{\sqrt{2}}(\chi+i\psi)$ as in \eqref{toy}, obtaining 
\begin{equation}
\begin{split}
\Lc = & \frac{1}{2}\Bigg[ \chi \Box\chi-\psi\Box\psi-m_i(\chi^2-\psi^2)+2n\chi\psi \\
&-3\pi G_{N}\left(\frac{x^{2}}{4}(\chi^{4} + \psi^{4}) +\frac{2y^{2}-x^{2}}{4}\chi^{2}\psi^{2} + x y \chi\psi(\chi^{2} - \psi^{2})\right)\Bigg] \, ,
\end{split}
\end{equation}
with $m= \frac{y^{2}-x^{2}+ 6 x H_{0}}{4}$ and $n = \frac{y(x-3H_{0})}{2}$.
To give a crude approximation on the lifetime of our vacuum, let us set to zero the quadratic term  $2n\chi\psi$.\footnote{This is actually a delicate point in our context. In fact setting  $n = 0$ is not just a simple choice of parameters. This would correspond to $y=0$ or $x=3H_{0}$, implying $Im(J)=0$. Therefore taking this specific limit do not correspond at all to a situation with two complex roots for the operator $\Fc$, but on the contrary corresponds to a situation with two real coincident roots. For simplicity let us however assume that we can neglect the quadratic term above.} Looking at the quartic part of the Lagrangian  we see that several interactions involving  phantoms are present. Following the rules above only a few decays are however allowed. In particular the kinematically permitted ones are $\psi\rightarrow \psi\psi\psi$, $\psi\rightarrow \psi\psi\chi$ and $\chi\rightarrow \chi\chi\psi$.
The decay rate for a  particle at rest for this  kind of processes can be written generically as
\begin{equation}
\Gamma = \frac{1}{m_{i}}\int  \frac{d^{3}p_{1}}{(2\pi)^{3}2 E_{1}}\int  \frac{d^{3}p_{2}}{(2\pi)^{2}2 E_{2}}\int  \frac{d^{3}p_{3}}{(2\pi)^{3}2 E_{3}}|\Mc|^{2}(2 \pi)^{4}\delta^{(4)}(p_{i}-p_{1}-p_{2}-p_{3}) \, .
\end{equation}
The prefactor is the mass (that is the same in all the processes here) of the particle that decay, $\Mc$ is the effective coupling constant of the quartic interaction corresponding to the process considered, in our case these are just tree level processes and the factor  $|\Mc|^{2}$ can be taken out of the integrals. The integrals over the momentum are up to a cutoff energy $\Lambda_{cutoff}$, as explained above. 
The coupling constant of all the processes corresponding to  quartic interaction in our Lagrangian is roughly just proportional to $G_{N}|\alpha|^{2} \sim|\alpha|^{2}/M_{P}^{2} $ and all the phantom and the canonical  field considered have the same mass squared, proportional to $|J|\sim|\alpha|^{2}$. On a purely dimensional analysis basis we can thus estimate the decay rate order of magnitude to be
\begin{equation}
\Gamma \sim \left(\frac{|\alpha|^2}{M^{2}_{P}} \right)^{2} \frac{\Lambda^{2}_{cutoff}}{m_{i}} \sim  \frac{|\alpha|^3}{M^{4}_{P}}\Lambda^{2}_{cutoff} \, .
\end{equation}
{If we consider the SFT origin of the class of model discussed, the order of magnitude of $\alpha$ is roughly of the same order  as the string mass, which is the inverse of the string length $\sqrt{\alpha'}$,
and  we can consider it to be related to $M_{P}$ as   $\alpha\sim z M_{P}$.} Here we have introduced a new parameter $z$, which is just the ratio of the string mass to the Planck mass \cite{Aref'eva:2008gj}. $z$ may vary in a wide range since experimentally any mass above, say, 10 TeV is not excluded as a string scale. The natural cutoff coming from SFT is  the string mass giving $\Lambda_{cutoff} \sim z M_{P}$. The decay mean time for the processes considered above expressed in the unit of Hubble time , $H^{-1}_{0} \sim 10^{60} M_{P}$, is of the order
\begin{equation}
H_{0}\tau_{decay} \sim \frac{10^{-60}}{z^{5}} \, .
\end{equation}

{It is obvious from the latter that the lifetime of the vacuum can be large and even roughly comparable with the age of the Universe. The price for this is the lightweight string. $z$, which determines this, can be in the range $10^{-15}\div1$,  since we do not expect strings lighter than 10 TeV from  experimental bounds and heavier than the Planck mass from the theoretical perspectives.}

Concerning the second question about the UV completion, the following idea might provide an answer  in the context of SFT, even if the solution is most likely not unique. The point is that in the class of model we are considering here, in contrast to canonical scalar field models, different vacua can easily have different number and nature of degrees of freedom. One may therefore expect that, apart from the vacuum corresponding to the present model, there is another vacuum which has no ghosts and is generally well behaved.
 To illustrate this point we can consider the following action
\begin{equation}
S=\int d^4x\left(\frac12\phi(\Box-m^2)\phi+v(\Gc(\Box)\phi)\right)\, ,
\label{ghosts-noghosts}
\end{equation}
where $\Gc$ is a non-local operator and $v$ describes some interaction potential. Now suppose that there are two minima with
different second derivatives of the function $v$ only. The kinetic operator would then look like
\begin{equation}
\Fc_k=\Box-m^2+v_k\Gc(\Box) \, ,
\label{ghosts-noghosts-K}
\end{equation}
where $v_k=v''(\Gc(0)\phi_k)$ and $\phi_k$ are  distinct vacuum solutions.
A specific function of the form $\Gc=\alpha(\Box-m^2)(e^{\beta\Box}-1)$ shows explicitly that in the vacuum with $v_k=1/\alpha$ there is only one root for the resulting function $\Fc_k$ while any other value of $v_k$ generates infinitely many roots and therefore ghosts.
Indeed, if $v_k=1/\alpha$ one has
\begin{equation*}
\Fc_k=(\Box-m^2)e^{\beta\Box}
\end{equation*}
and this kinetic operator corresponds to one scalar degree of freedom.
Even though this is not a genuine SFT derived action, it clearly illustrates how the model can be UV completed.

The bottom line is that the parameters of the model can be such that an effect like the phantom divide crossing is observable and is extended  over an appreciable lapse of time, but the eventual stabilization of the model is most likely achieved as  a transition to a really stable vacuum and the discussion in the second half of this section describes how this can happen.
Here it is important to mention that, as it is discussed in \cite{Cline:2003gs,Woodard:2006nt}, the ghost instability time scale can be finite only if the cutoff is Lorentz violating as in our consideration where we have used a preferred frame. This is not what one expects from the viewpoint of SFT. It is however not excluded that the false vacuum of SFT which contains ghosts could be Lorentz-violating. This question definitely requires better understanding.

\section{Conclusions}

In this paper we have studied a class of models which describe the late time
evolution in the dynamics of  scalar excitations in  SFT. This class of
models admits simple analytic solutions when minimally coupled to gravity. The main distinctive
feature is the presence of non-locality arising from the presence of an operator
$\Fc(\Box)$ in the action. This non-locality can be translated in a (possibly
infinite) set of local fields with masses squared $J_i$. These masses squared may be complex.
The corresponding complex fields enter the Lagrangian in an unusual way, giving rise to a new structure.
The main question we have addressed  is 
the influence of such complex fields on the evolution of  linear perturbations of the system.
In the meantime, we have shown explicitly that these complex fields give a
solution describing the crossing of the phantom divide. Therefore stability (or
instability) under  linear perturbations is essential to understand whether our model
can (or cannot) be used as a viable model for the explanation of such a
phenomenon.

In order to have  good background solutions we must require  that the parameters $\alpha_{i}$ of the model  satisfy 
 $\text{Re}(\alpha_i)<0$.
Then the solutions for the scalar fields do not blow up and the Hubble parameter tends to a constant.
The  most important result is the fact that, having such
well behaved solutions,  the perturbations related to complex $\alpha_i$ eventually vanish
for any values of the other parameters (the comoving wavenumber $k$, $\text{Im}(\alpha_i)$, the initial data
for the background solutions and perturbations).
However, we notice that for some range of the parameters  the perturbations do exceed
the perturbative regime, making the question of validity of the linear
approximation unclear. This range of parameters certainly deserves a more extensive
 analysis including the second order perturbations. 

From our analysis it turns out  that in a general situation where real and
complex $J_i$ are
involved, the perturbative spectrum  is determined just by 
the real ones for a wide range of the parameters. Indeed, real values of
$J_i$ would generate ordinary real scalar fields. Cosmological perturbations with
such scalar fields are well  understood  and are known to have growing or
vanishing modes  mainly depending on the values of the  wavenumber $k$.
This leads us to state that, since perturbations related to 
complex $J_i$ almost always vanish,  the observable spectrum of the model 
is determined solely by  real $J_i$.
Thus the crossing of the phantom divide,
 realized in the case of complex $J_i$,  is perturbatively stable
and perturbations do not grow even though the phantom phase is
reached.

The presence of ghost-like excitations however is alarming. We have computed an estimate for the vacuum decay rate due to ghosts in the context of our simple model. Admitting that the string mass is not necessarily of the order of Planck mass, one finds a quite large range of values for the lifetime. This can make the described effect of the phantom divide crossing observable. Here we want to stress again that, in SFT based models, the appearance of ghosts is most likely  an indication of the fact that there exists another vacuum to which the model evolves. It is exactly the existence of a proper vacuum that  is  suggested as a way to make the model UV-complete. Even though there can be parameters  such a that it is suitable for the present day observations, the model can have relevance for early stages of the Universe evolution, when effects like the phantom divide crossing were also possible.

To summarize we conclude that models of type \eqref{action_model_localexact}
provide interesting  results from the point of view of cosmology.
Apart from the phantom divide crossing, the appearance of  non-gaussianity
during the inflation is another fact which makes the class of models considered
in the present paper promising from the perspective of future cosmological
studies. The results of this paper are generally applicable and one can state that 
perturbations in similar systems would not be triggered by new type of fields and quantum effects can be suppressed for a long time according to the vacuum decay rate estimate.
Therefore, this new class of models is believed to provide new successful ways
of resolving some of the cosmological questions. Further aspects, like coupling
to
other cosmic fluids (CDM, radiation, etc), construction of other potentials, consideration of 
thermodynamical properties\footnote{See, for instance, \cite{kapusta} about
interesting
and promising results concerning the thermodynamics of  $p$-adic strings.}
as well as better understanding of the UV-completion mechanism certainly  
deserve  a deeper study.


\acknowledgments
The authors are grateful to I.Ya.~Aref'eva, 
F.~Bezrukov, B.~Craps, B.~Dragovich, G.~Dvali,
V.~Mu\-kha\-nov and  S.Yu.~Vernov for useful comments and stimulating
discussions. This work  is supported in part by the Belgian Federal Science Policy Office
 through the Interuniversity Attraction Pole P6/11, and in part by the FWO-Vlaanderen through the project G.0114.10N.
A.K. is supported in part by RFBR grant 08-01-00798 and state contract of
Russian Federal Agency for Science and Innovations 02.740.11.5057.


\end{document}